\documentclass{iau}
\usepackage{graphicx}

\title[Asteroseismology of fast-rotating stars] 
{Asteroseismology of fast-rotating stars :\\ the example of $\alpha$ Ophiuchi}

\author[Mirouh G.M., Reese D.R., Espinosa Lara F., Ballot J. \& Rieutord M.]   
{Giovanni M. Mirouh$^{1,2}$, Daniel R. Reese$^3$, Francisco Espinosa Lara$^{1,2}$, J{\'e}r{\^o}me Ballot$^{1,2}$ \& Michel Rieutord$^{1,2}$ }

\affiliation{$^1$ Universit{\'e} de Toulouse, UPS-OMP, IRAP, Toulouse, France \\[\affilskip]
$^2$ CNRS, IRAP, 14 avenue Edouard Belin, 31400 Toulouse, France \\[\affilskip]
$^3$ Institut d'Astrophysique et G{\'e}ophysique de l'Universit{\'e} de Li{\`e}ge, All{\'e}e du 6 Ao{\^u}t 17, 4000 Li{\`e}ge, Belgium}

\pubyear{2013}
\volume{301}  
\pagerange{119--120}
\jname{Precision asteroseismology}
\editors{W. Chaplin, J. Guzik, G. Handler \& A. Pigulski, eds.}
\begin{document}

\maketitle

\begin{abstract}
Many early-type stars have been measured with high angular velocities.
In such stars, mode identification is difficult as the effects of
fast and differential rotation are not well known.
Using fundamental parameters measured by interferometry, the
ESTER structure code and the TOP oscillation code, we investigate the
oscillation spectrum of $\alpha$ Ophiuchi, for which observations by
the MOST satellite found 57 oscillations frequencies. Results do not
show a clear identification of the modes and highlight the difficulties of
asteroseismology for such stars with a very complex oscillation
spectrum.

\end{abstract}

\firstsection 

\section{Introduction}

Intermediate- and high-mass stars are usually fast rotators. In some
of these A-, B- and O-type stars, the $\kappa$-mechanism excites
eigenmodes. Because of the centrifugal flattening of the star and the
strength of the Coriolis acceleration, the oscillation spectrum is
much more complex than that of non-rotating stars. Interpretation
of observed frequencies of these stars requires two-dimensional
models and their oscillation spectrum computed in a non-perturbative
way. Following the pioneering work of \cite{deupree_etal12}, we want 
to identify the observed modes of the fast-rotating star $\alpha$ 
Ophiuchi in order to better
constrain its fundamental parameters. Compared to the previous work of
Deupree, we improve the numerical resolution and get spectrally
converged eigenfunctions that allow a reliable computation of
visibilities and damping rates of the modes.

\section{Models}

  \begin{figure}[t]
    \includegraphics[width=0.31\textwidth]{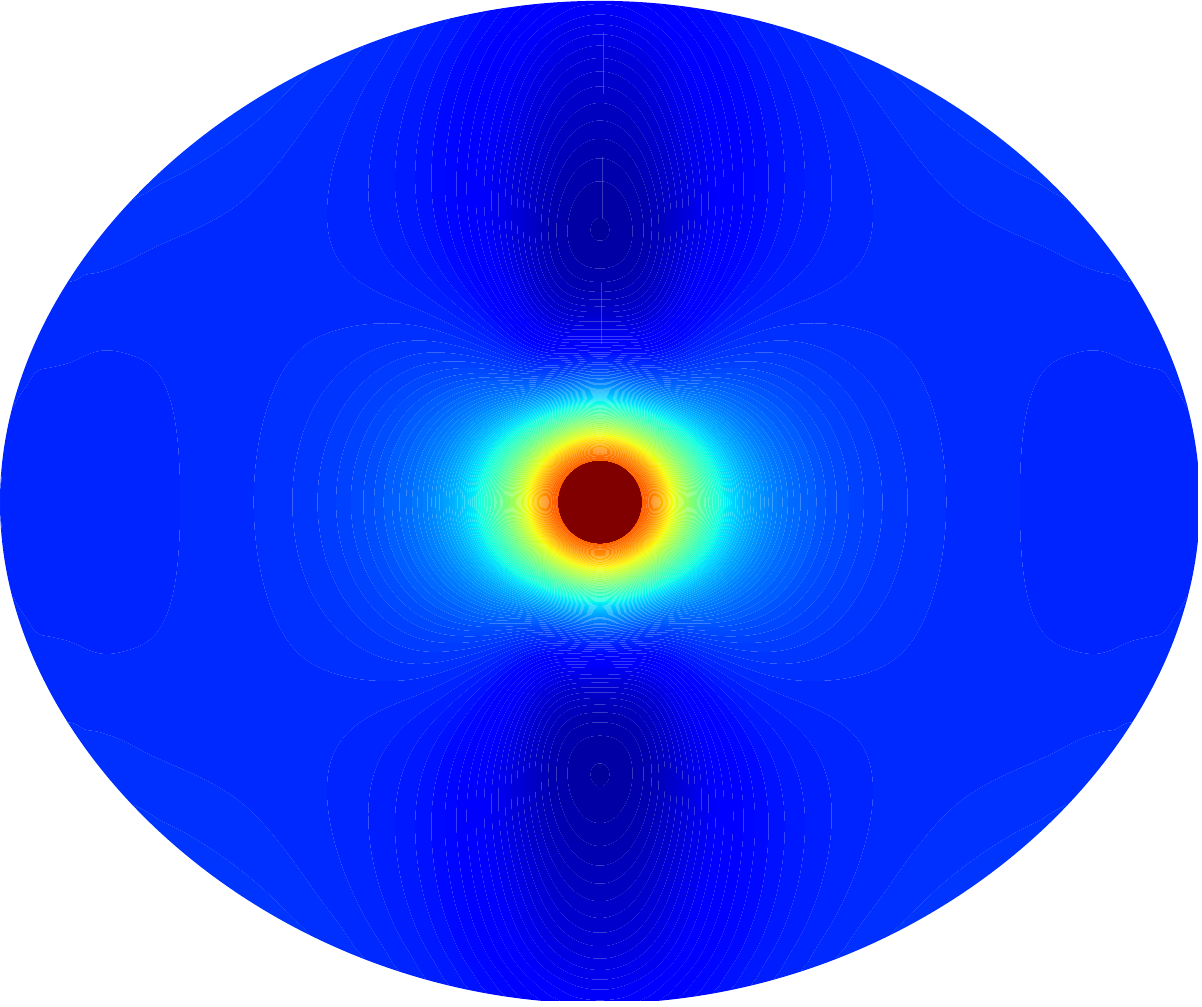}     \hfill
    \includegraphics[width=0.31\textwidth]{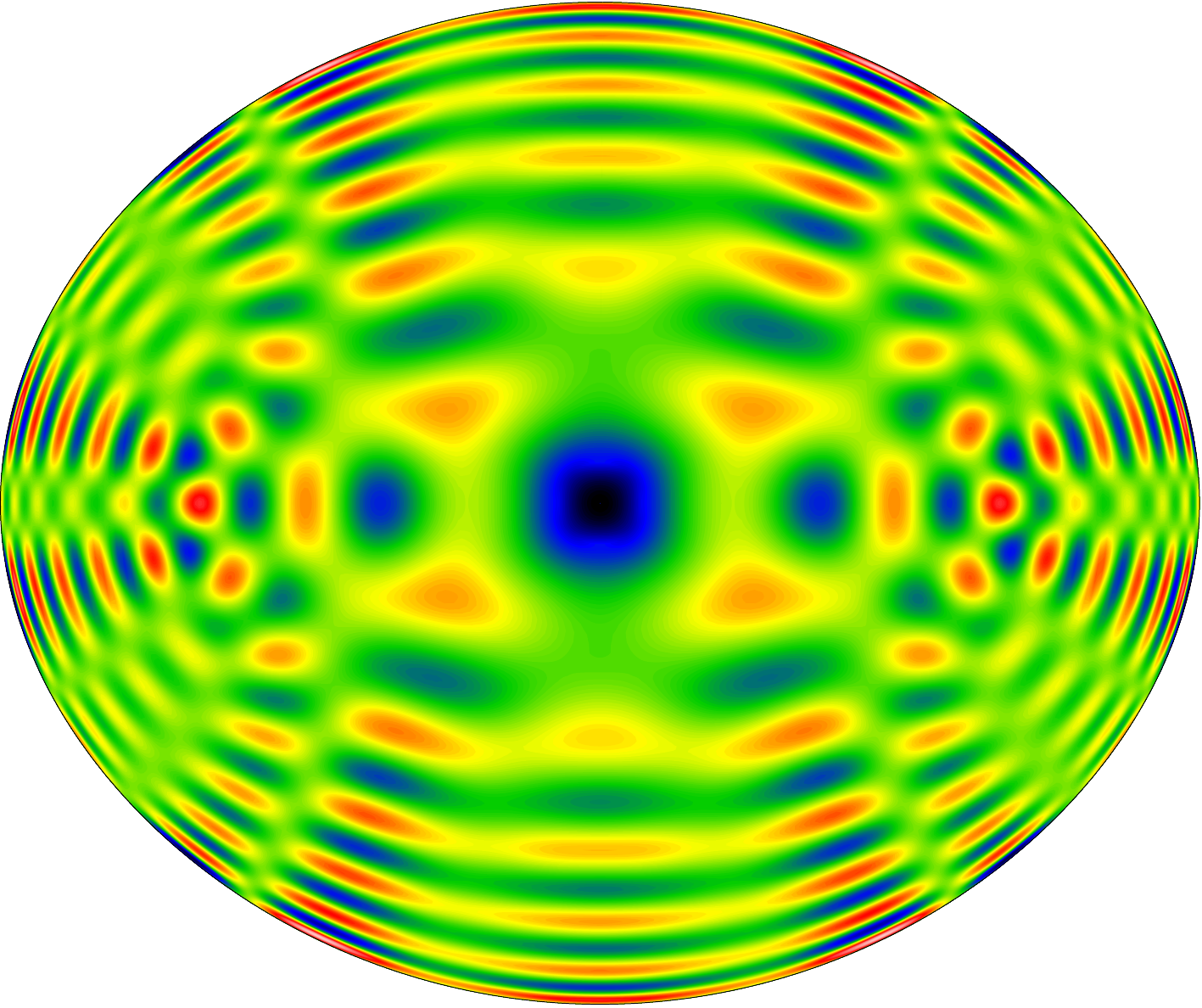}	    \hfill
    \includegraphics[width=0.31\textwidth]{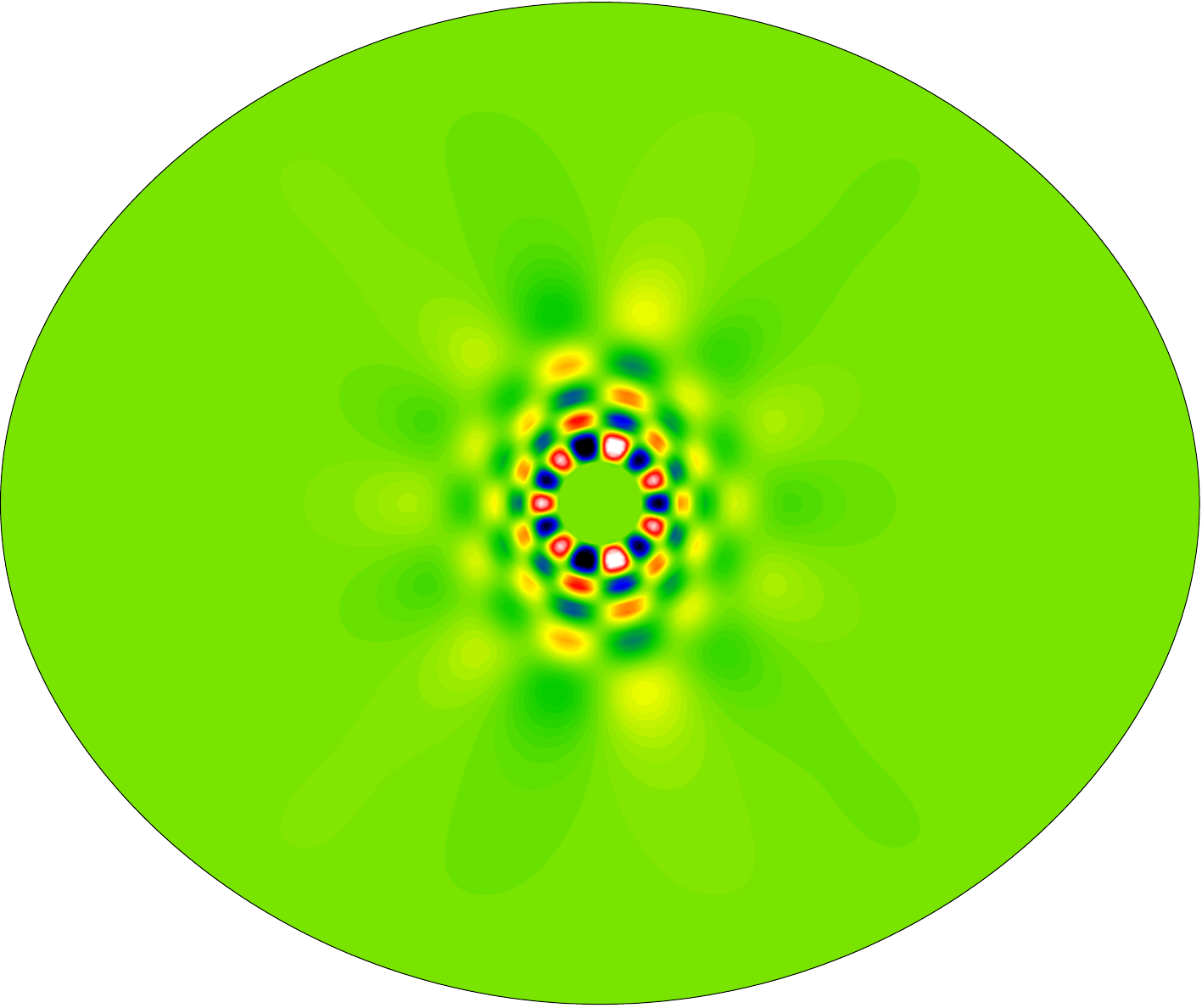}
    \caption{From left to right: Rotation profile in $\alpha$ Oph (fast
core, slow envelope), an $m=0$ acoustic island-mode, an $m=-1$ g-mode.}
    \label{fig:star}
  \end{figure}

We compute 2D models for the fast-rotating A-type star Rasalhague
($\alpha$ Ophiuchi) with the ESTER code (\cite{REL13}), in which
differential rotation is calculated self-consistently. Its surface
equatorial rotation velocity of 240 km$\cdot$s$^{-1}$ imposes a flatness of
0.168. Its equatorial and polar radii have been derived from interferometry
(\cite{zhao_etal09}) and the stellar mass is constrained by an orbiting
companion (\cite{hinkley_etal11}).  Besides this, 57 oscillation frequencies have
been measured by photometry (\cite{monnier_etal10}). To reproduce the fitted radii, 
luminosity, polar and equatorial temperatures,our model uses $M= 2.22 M_\odot$, 
$\Omega = 0.62 \Omega_{\rm K} = 1.65$ c/d, mass fractions 
 $Z = 0.02$ and $X=0.7$ in the envelope, $X_c = 0.26$ 
in the core (see \cite{ELR13}).  The eigenvalue problem of
adiabatic oscillations is solved with the TOP code (\cite{reese_etal09})
for modes with azimutal orders $-4\leq m \leq 4$, in the range of frequencies in which
the p modes are thought to be observed (see Figure \ref{fig:star}).

\section{Comparisons between computed and observed modes}

    \begin{figure}[h!]
     \centering
      \includegraphics[width=0.8\textwidth]{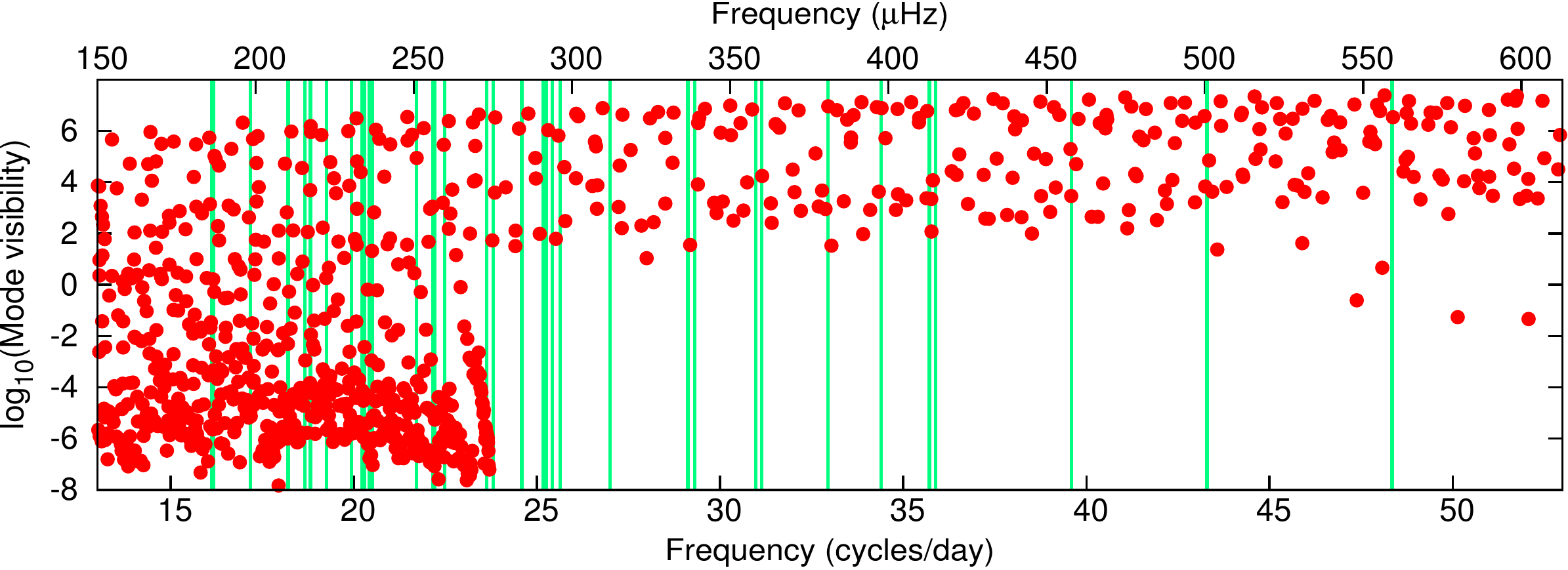}
      \caption{Visibilities for axisymmetric ($m=0$) modes. The vertical
      lines indicate the frequencies measured with MOST (Monnier et
      al. 2010).}
      \label{figvis}
    \end{figure}

To select the modes that might be seen from Earth, we compute the mode
visibilities, following \cite{reese_etal13} and the thermal dissipation
rates using the quasi-adiabatic approximation (\cite{unno_etal89}). In
the domain that we investigated, our model yields only linearly stable
modes. These are g modes and p modes modified by rotation. g modes are
the least-damped with large amplitude at the base of the envelope. Their
visibility is much less than that of the p modes that exist in this interval,
as their amplitude is evanescent at the surface.
However, p modes are much more damped. As shown in
Fig.~\ref{figvis}, where we clearly distinguish the set of g modes on
the low-frequency side, each observed frequency corresponds to several
eigenmodes of the model.  The damping rates do not seem to be able to
lift this degeneracy of the matching. It may well be that our models are
at the moment too simple, notably because of the chemical homogeneity
of the envelope. Progress therefore calls for more realistic models that
include the distribution of chemical elements resulting from time evolution.

\end{document}